\documentclass[10pt]{bmc_article}
\usepackage{subfigure}
\usepackage{float}
\usepackage{amssymb}
\usepackage{amsmath}
\usepackage{amsbsy}
\usepackage{mario-def}
\usepackage{booktabs}
\usepackage{multirow}
\usepackage{color}

\usepackage{graphicx}
\usepackage[space]{grffile}
\usepackage[bookmarks=false, colorlinks,citecolor=black,filecolor=black,linkcolor=black,urlcolor=black]{hyperref}
\usepackage{caption}
\usepackage{cite} 
\usepackage{url}  
\usepackage{ifthen}  
\usepackage{multicol}   
\usepackage[utf8]{inputenc} 
\newboolean{publ}

\newtheorem{theorem}{Theorem}

\newenvironment{bmcformat}{\sloppy\setboolean{publ}{false}}{\sloppy}

\begin{document}
\begin{bmcformat}


\title{Learning a peptide-protein binding affinity predictor with kernel ridge regression}
 

\author{Sébastien Giguère\correspondingauthor$^1$
         \email{Sébastien Giguère\correspondingauthor - sebastien.giguere.8@ulaval.ca}\and
         Mario Marchand$^1$
         \email{Mario Marchand - mario.marchand@ift.ulaval.ca}\and
         François Laviolette$^1$
         \email{François Laviolette - francois.laviolette@ift.ulaval.ca}\and
         Alexandre Drouin$^1$
         \email{Alexandre Drouin - alexandre.drouin.8@ulaval.ca}\and
         Jacques Corbeil$^2$
         \email{Jacques Corbeil - jacques.corbeil@crchul.ulaval.ca}
      }


\address{%
    \iid(1)Department of Computer Science and Software Engineering, Université Laval, Québec, Canada\\
    \iid(2)Department of Molecular Medicine, Université Laval, Québec, Canada
}%

\maketitle

\vspace{-0.5cm}
\begin{abstract}

\vspace{-0.5cm}
\subsection*{Background}

The cellular function of a vast majority of proteins is performed through physical interactions with other biomolecules, which, most of the time, are other proteins. Peptides represent templates of choice for mimicking a secondary structure in order to modulate protein-protein interaction. They are thus an interesting class of therapeutics since they also display strong activity, high selectivity, low toxicity and few drug-drug interactions.
Furthermore, predicting which peptides would bind to which MHC alleles would be of tremendous benefit to improve vaccine based therapy and possibly generating antibodies with greater affinity.
Modern computational methods have the potential to accelerate and lower the cost of drug and vaccine discovery by selecting potential compounds for testing in silico prior to biological validation.
\vspace{-0.25cm}
\subsection*{Results}
We propose a specialized string kernel for small bio-molecules, peptides and pseudo-sequences of binding interfaces.
The kernel incorporates physico-chemical properties of amino acids and elegantly generalize eight kernels, such as the Oligo, the Weighted Degree, the Blended Spectrum, and the Radial Basis Function.
We provide a low complexity dynamic programming algorithm for the exact computation of the kernel and a linear time algorithm for it's approximation.
Combined with kernel ridge regression and SupCK, a novel binding pocket kernel, the proposed kernel yields biologically relevant and good prediction accuracy on the PepX database.
For the first time, a machine learning predictor is capable of accurately predicting the binding affinity of any peptide to any protein.
The method was also applied to both single-target and pan-specific Major Histocompatibility Complex class II benchmark datasets and three Quantitative Structure Affinity Model benchmark datasets.
\vspace{-0.25cm}
\subsection*{Conclusion}
On all benchmarks, our method significantly (p-value $\le 0.057$) outperforms the current state-of-the-art methods at predicting peptide-protein binding affinities.
The proposed approach is flexible and can be applied to predict any quantitative biological activity.
Moreover, generating reliable peptide-protein binding affinities will also improve system biology modelling of interaction pathways.
Lastly, the method should be of value to a large segment of the research community with the potential to accelerate peptide-based drug and vaccine development.

\end{abstract}

\ifthenelse{\boolean{publ}}{\begin{multicols}{2}}{}



\section{Background}

The cellular function of a vast majority of proteins is performed through physical interactions with other biomolecules, which, most of the time, are other proteins. Indeed, essentially all of the known cellular and biological processes depend, at some level, on protein-protein interactions (PPI)~\cite{Toogood2002Inhibition, Albert2005Scalefree}. Therefore, the controlled interference of PPI with chemical compounds provides tremendous potential for the discovery of novel molecular tools to improve our understanding of biochemical pathways as well as the development of new therapeutic agents~\cite{Wells2007Reaching, Domling2008Small}. 

Considering the nature of the interaction surface, protein secondary structures are essential for binding specifically to protein interaction domains. Peptides also represent templates of choice for mimicking a secondary structure in order to modulate protein-protein interactions \cite{Costantino2006Privileged, PerezDeVegaModulation}. Furthermore, they are a very interesting class of therapeutics since they display strong activity, high selectivity, low toxicity and fewer drug-drug interactions. They can also serve as investigative tools to gain insight into the role of a protein, by binding to distinct regulatory regions to inhibit specific functions.

Yearly, large sums of monies are invested in the process of finding druggable targets and identifying compounds with medicinal utility. The widespread use of combinatorial chemistry and high-throughput screening in the pharmaceutical and biotechnology industries implies that millions of compounds can be tested for biological activity.
However, screening large chemical libraries generates significant rates of both false positives and negatives.
The process is expensive and faces a number of challenges in testing candidate drugs and validating the hits, all of which must be done efficiently to reduce costs and time.
Computational methods with reasonable predictive power can now be envisaged to accelerate the process providing an increase in productivity at a reduced cost.

Furthermore, peptides ranging from 8 to 12 AA represent the recognition unit for the immune system and in particular the MHC (Major Hiscompatibility Complex).
Being capable of predicting which peptides would bind to which MHC alleles would be of tremendous benefit to improve vaccine based therapy, possibly generating antibodies with greater affinity that could yield an improved immune response.
Moreover, simply having data on the binding affinity of peptides and proteins could readily assist system biology modelling of interaction pathways.

The ultimate goal is to build a predictor of the highest binding affinity peptides.
If one had a fast and accurate peptide-protein binding affinity predictor, one could easily predict the affinity of huge sets of peptides and select the best candidates, or use stochastic optimization methods if the set of peptides were too large.
The task would be facilitated if one could build such a fast and accurate predictor.
This paper provides a step in this direction with the use of a machine learning algorithm based on kernel methods and a novel kernel.

Traditional machine learning approaches used to focus on binary classification of compounds (binding, non-binding) \cite{Jacob2008Virtual}.
Non-binding compounds are rarely known and available quantitative binding affinity information is lost during training, a major obstacle to binary classification.
New databases, such as the PepX database, contain binding affinities between peptides and a large group of protein families. 
The first part of this paper presents a general method for learning a binding affinity predictor between any peptide and any protein, a novel machine learning contribution to biology.

The Immune Epitope Database (IEDB) \cite{Peters2005} contains a large number of binding affinities between peptides and Major Histocompatibility Complex (MHC) alleles.
Predicting methods for MHC class I alleles have already had great success \cite{Zhang2011Toward}.
The simpler binding interface of MHC-I molecules makes the learning problem significantly easier than for MHC-II molecules. 
Allele specific (single-target) methods for MHC class II alleles have also reasonable accuracy, despite requiring a large number of training data for every allele in order to achieve reasonable accuracy \cite{Zhang2011Toward}.
Pan-specific (multi-target) methods, such as MultiRTA\cite{Bordner2010MultiRTA} and NetMHCIIpan-2.0\cite{Nielsen2010}, were designed in order to overcome this issue.
These methods can predict, with reasonable accuracy, the binding affinity of a peptide to any MHC allele, even if this allele has no known peptide binders.

We propose a new machine learning approach based on kernel methods~\cite{sc-04} capable of both single-target and multi-target (pan-specific) prediction.
We searched for kernels that encode relevant binding information for both proteins and peptides.
Therefore, we propose a new kernel, the GS kernel, that generalizes most of kernels currently used in this setting (RBF~\cite{sc-04}, Blended spectrum~\cite{sc-04}, Oligo~\cite{Meinicke2004Oligo}, Weighted Degree~\cite{RS2004}, ...).
The GS kernel is shown to be a suitable similarity measure between peptides and pseudo-sequences of MHC-II binding interfaces.

For the machine learning algorithm itself, we show that kernel ridge regression~\cite{sc-04} (KRR) is generally preferable to the support vector regression (SVR) algorithm~\cite{ss-98} because KRR has less hyperparameters to tune than SVR, thus making the learning time smaller.
The regression score that we obtain on PepX is competitive with the ones that we obtain on data sets containing peptides binding to a single protein, even if the former task is, in theory, much more difficult.
For the peptide-MHC binding problem, comparison on benchmark datasets show that our algorithm surpasses NetMHCIIpan-2.0\cite{Nielsen2010}, the current state-of-the-art method.
Indeed, in the most difficult pan-specific case (when the algorithm is trained on all alleles except the allele used for testing), our algorithm performs better than the state of the art in most cases.
Finally, our method outperform SVR on three quantitative structure affinity model (QSAM) single-target predictions benchmarks~\cite{Zhou2010Gaussian}.
We thus propose a machine learning approach to immunology and a novel string kernel which have shown to yield impressive results on benchmark datasets for various biological problems.

\section{Methods}

\subsection{Statistical machine learning and kernel ridge regression in our context}

\medskip
Given a set of training examples (or cases), the task of a learning algorithm is to build a accurate predictor. In this paper, each example will be of the form $((\xb,\yb),e)$, where $\xb$ represents a peptide, $\yb$ represents a protein, and $e$ is a real number representing the binding energy (or the binding affinity) between the peptide $\xb$ and the protein $\yb$. A multi-target predictor is a function $h$ that returns an output $h(\xb,\yb)$ when given any input~$(\xb,\yb)$.  In our setting, the output $h(\xb,\yb)$ is a real number estimate of the ``true'' binding energy (or the binding affinity) $e$ between $\xb$ and $\yb$.  The predictor~$h$ is accurate on example $((\xb,\yb),e)$ if the predicted output $h(\xb,\yb)$ is very similar to the real output~$e$. A predictor is good when it is accurate on most future examples unseen during training.

With kernel methods, each input $(\xb,\yb)$ is mapped to a \emph{feature vector} $\phib (\xb,\yb) = (\phi_1(\xb,\yb), \phi_2(\xb,\yb),\ldots,\phi_d(\xb,\yb))$ of large dimensionality $d$. 
Moreover, the predictor is represented by a real-valued weight vector $\mathbf{w}$ that lies in the space of feature vectors. Given an arbitrary input $(\xb,\yb)$, the output of the predictor $h_{\mathbf{w}}$ is  given by the scalar product $$h_{\mathbf{w}}(\xb,\yb) = \mathbf{w}\cdot\pmb \phi(\xb,\yb)\eqdef \sum_{i=1}^d w_i\phi_i(\xb,\yb)\,.$$ 

The lost incurred by predicting a binding energy $h_\wb(\xb,\yb)$ on input $(\xb,\yb)$, when the true binding energy is $e$, is measured by a \emph{loss function} $\ell(\wb,(\xb,\yb),e)$. As is usual in regression, we will use the quadratic loss function $$\ell(\wb,(\xb,\yb),e) = (e-\wb\cdot\phib(\xb,\yb))^2\, .$$ 

The fundamental assumption in machine learning is that each example $((x,y),e)$ is drawn according to some unknown distribution $D$. Then the task of the learning algorithm is to find the predictor $h_\wb$ having the smallest possible \emph{risk} $R(h_\wb)$ defined as the expected loss $$R(h_\wb)\eqdef\esp{((\xb,\yb),e)\sim D}\ell(\wb,(\xb,\yb),e)\, .$$ 

However, the learning algorithm does not have access to $D$. Instead, it has access to to a training set $ S \eqdef \{ ((\xb_1, \yb_1), e_1), ((\xb_2,\yb_2), e_2),\ldots, ((\xb_m,\yb_m), e_m)\}$ of $m$ examples where each example $((\xb_i, \yb_i), e_i)$ is assumed to be generated independently according to the same (but unknown) distribution $D$. Modern statistical learning theory~\cite{sc-04,ss-02} tells us that the predictor $h_\wb$ minimizing the \emph{ridge regression cost function} $F(S,\wb)$ will have a small risk $R(h_\wb)$ whenever the obtained value of $F(S,\wb)$ is small. Here, $F(S,\wb)$ is defined as 
\begin{equation*}
 F(S,\wb)\ \eqdef\ \|\mathbf{w}\|^2 + C \sum_{i=1}^m\ell(\wb,(\xb_i,\yb_i),e_i) = \ \|\mathbf{w}\|^2 + C \sum_{i=1}^m (e_i-\wb\cdot\phib(\xb_i,\yb_i))^2\, , 
\end{equation*}
for some suitably-chosen constant $C>0$. The first term of $F(S,\wb)$, $\|\wb\|^2\eqdef \wb\cdot\wb$, which is the squared Euclidean norm of $\wb$, is called a \emph{regularizer} and it penalizes predictors having a large norm (complex predictors). The second term measures the accuracy of the predictor on the training data. Consequently, the parameter $C$ controls the complexity-accuracy trade-off. Its value is usually determined by measuring the accuracy of the predictor on a separate (``hold-out'') part of the data that was not used for training, or by more elaborate sampling methods such as cross-validation.

The \emph{representer theorem}~\cite{sc-04,ss-02} tells us that the predictor $\mathbf{w}^*$ that minimizes $F(S,\wb)$ lies in the linear subspace span by the training examples. In other words, we can write $$\mathbf{w}^* = \sum_{i=1}^m\alpha_i\pmb \phi(\xb_i,\yb_i)\, ,$$  where the coefficients $\alpha_i$ are called the \emph{dual} variables and provide collectively the dual representation of the predictor. This change of representation makes the cost function dependent on $\pmb \phi(\xb_i)$ only via the scalar product $\pmb \phi(\xb_i,\yb_i) \cdot \pmb \phi(\xb_j,\yb_j)\eqdef k((\xb_i,\yb_i), (\xb_j,\yb_j))$ for each pair of examples. The function $k$ is called a \emph{kernel} and has the property of being efficiently computable for many feature maps~$\pmb \phi$, even if  the feature space induced by $\pmb \phi$ has an extremely large dimensionality. By using $k$ instead of $\pmb \phi$, we can construct linear predictors in feature spaces of extremely large dimensionality with a running time that scales only with the size of the training data (with no dependence on the dimensionality of $\pmb \phi$). This fundamental property is also known as the \emph{kernel trick}~\cite{sc-04,ss-02}. It is important to point out that, since a kernel corresponds to a scalar product in a feature space, it can be considered as a similarity measure. A large (positve) value of the kernel normally implies that the corresponding feature vectors point in the same direction (similar), although a value close to zero normally implies that the two feature vectors are mostly orthogonal (dissimilar). 

We restrict ourselves to joint feature maps having the form $\phib(\xb,\yb) = \phib_\Xcal(\xb)\otimes \phib_\Ycal(\yb)$ where $\otimes$ denotes the tensor product. The tensor product between two vectors $\ab =(a_1,\ldots,a_n)$ and $\bb=(b_1,\ldots,b_m)$ denotes the vector $\ab\otimes\bb = (a_1b_1, a_1b_2,\ldots,a_nb_m)$ of all the $nm$ products between the components of $\ab$ and $\bb$. 
The tensor product allows us to express the joint feature space $\phib$ in terms of the feature space $\phib_\Xcal$ of the peptides and the feature space $\phib_\Ycal$ of the proteins. If we now define the peptide kernel $k_\Xcal$ by $k_\Xcal(\xb,\xb')\eqdef\phib_\Xcal(\xb)\cdot\phib_\Xcal(\xb')$, and the protein kernel $k_\Ycal$ by $k_\Ycal(\yb,\yb')\eqdef\phib_\Ycal(\yb)\cdot\phib_\Ycal(\yb')$, the joint kernel $k$ simply decomposes as the product of $k_\Xcal$ and $k_\Ycal$ as 
\begin{eqnarray*}
k((\xb,\yb),(\xb',\yb')) &\eqdef&  \phib(\xb,\yb)\cdot\phib(\xb',\yb')\\ 
&=& \phib_\Xcal(\xb)\otimes\phib_\Ycal(\yb)\cdot\phib_\Xcal(\xb')\otimes\phib_\Ycal(\yb')\\ 
&=& (\phib_\Xcal(\xb)\cdot\phib_\Xcal(\xb'))(\phib_\Ycal(\yb)\cdot\phib_\Ycal(\yb'))\\ 
&\eqdef& k_\Xcal(\xb,\xb') k_\Ycal(\yb,\yb')\, . 
\end{eqnarray*}
Consequently, from the representer theorem we can write the multi-target predictor as
 $$h_{\wb^*}(\xb,\yb) = \wb^*\cdot\phib(\xb,\yb) =  \wb^*\cdot(\phib_\Xcal(\xb)\otimes \phib_\Ycal(\yb)) = \sum_{i=1}^m \al_i k_\Xcal(\xb_i, \xb)k_\Ycal(\yb_i, \yb)\, .$$  

In the case of the quadratic loss $\ell(\wb,(\xb,\yb),e) = (e-\wb\cdot\phi(\xb,\yb))^2$, $F(S,\wb)$ is a strongly convex function of $\wb$ for any strictly positive $C$. In that case, there exists a single local minimum which coincides with the global minimum. This single minimum is given by the point $\wb^*$ where the gradient $\partial F(S,\wb)/\partial\wb$ vanishes. For the quadratic loss, this solution $\wb^*$ is given by  
\begin{equation}\label{eq:RRsolution} 
\alb = \LP\Kb + \frac{1}{C}\Ib\RP^{-1}\eb\, ,
\end{equation}
where $\alb\eqdef(\al_1,\ldots,\al_m)$, $\eb\eqdef(e_1,\ldots,e_m)$, $\Kb$ denotes the Gram matrix of kernel values $K_{i,j} = k_\Xcal(\xb_i,\xb_j) k_\Ycal(\yb_i,\yb_j)$, and $\Ib$ denotes de $m\times m$ identity matrix. Hence, the learning algorithm for kernel ridge regression just consists at solving Equation~\eqref{eq:RRsolution}. Note that for a symmetric positive semi-definite kernel matrix $\Kb$, the inverse of $\Kb + \Ib/C$ always exists for any finite value of $C> 0$. Note also that the inverse of an $m\times m$ matrix is obtained in $O(m^3)$ time with the Gaussian-elimination method and in $O(m^{2.376})$ time with the Coppersmith-Winograd algorithm.  
 
 \bigskip
Finally, we will also consider the single protein target case where only one protein $y$ is considered. In this case, the predictor $h_\wb$ predicts the binding energy from a feature vector $\phib_\Xcal$ constructed only from the peptide. Hence, the predicted binding energy for peptide $\xb$ is now given $\wb\cdot\phib_\Xcal(\xb)$. So, in this single protein target case, the cost function to minimize is still given by $F(S,\wb)$ but with $\phib(\xb,\yb)= \phib_\Xcal(\xb)$.  Consequently, in this case, the solution is still given by Equation~\eqref{eq:RRsolution} but with a kernel matrix $\Kb$ given by $K_{i,j} = k_\Xcal(\xb_i,\xb_j)$.
The single-target predictor is thus given by
 $$h_{\wb^*}(\xb) =  \wb^*\cdot\phib_\Xcal(\xb) = \sum_{i=1}^m \al_i k_\Xcal(\xb_i, \xb)\, .$$

Kernel methods have been extremely successful within the last decade, but the choice of the kernel is critical for obtaining good predictors. Hence, confronted with a new application, we must be prepared to design an appropriate kernel. The next subsections show how we have designed and chosen both peptide and protein kernels. 

\subsection{A generic string (GS) kernel for small bio-molecule strings}\label{sec:GSkernel}

String kernels for bio-molecules have been applied with success in bioinformatics and computational biology.
Kernels for large bio-molecules, such as the local-alignment kernel \cite{Saigo2004} have been well studied and applied with success to problems such as protein homology detection. However, we observed that these kernels perform rather poorly on smaller compounds (data not shown).
Consequently, kernels designed for smaller bio-molecules like peptides and pseudo sequences have recently been proposed. Some of these kernels \cite{Meinicke2004Oligo} exploit sub-string position uncertainty while others \cite{twkr-10} use physicochemical properties of amino acids. We present a kernel for peptides that exploits both of these properties in a unified manner.

The proposed kernel, which we call the generic string (GS) kernel, is a similarity measure defined for any pair $(\xb,\xb')$ of strings of amino acids. Let $\Sigma$ be the set of all amino acids. Then, given any string $\xb$ of amino acids (e.g., a peptide), let $|\xb|$ denote the length of string $\xb$, as measured by the number of amino acids in $\xb$.
The positions of amino acids in $\xb$ are numbered from $1$ to $|\xb|$. 
In other words, $\xb=x_1,x_2,\ldots,x_{|\xb|}$ with all $x_i\in\Sigma$.

Now, let $\psib:\Sigma\longrightarrow \mathbb{R}^d$ be an encoding function such that for each amino acid~$a$, 
\begin{eqnarray}\label{eq:psib}
\psib\mbox{\small $(a)$}=(\psi_1\mbox{\small $(a)$},\psi_2\mbox{\small $(a)$},\ldots\psi_d\mbox{\small $(a)$}) 
\end{eqnarray}
is a vector where each component $\psi_i(a)$ encodes one of the $d$ properties (possibly physicochemical) of amino acid $a$.
In a similar way, we define $\psib^l: \Sigma^l\longrightarrow\mathbb{R}^{dl}$ as an encoding function for strings of length $l$.
Thus, $\psib^l(\ab)$ encodes all $l$ amino acids of $\ab$ concatenning $l$ vectors, each of $d$ components:

\begin{equation}\label{eq:psitol}
\psib^l(a_1,a_2,..,a_l) \eqdef (\psib(a_1), \psib(a_2), \ldots, \psib(a_l))
\end{equation}

Let $L\geq 1$ be a maximum length for substring comparison. We define the generic string (GS) kernel as the following similarity function over any pair $(\xb,\xb')$ of strings: 
\begin{equation}\label{GSkernel}
GS(\xb,\xb', L, \sigma_p, \sigma_c) \eqdef \sum_{l=1}^{L}\sum_{i=0}^{|x|-l} \sum_{j=0}^{|x'|-l}\,  e^{\left(\mbox{\large $\frac{-(i - j)^2}{2\sigma_p^2}$}\right)} \ e^{\left(\mbox{\large $\frac{-\parallel \psib^l(x_{i+1},..,x_{i+l}) \,-\, \psib^l(x'_{j+1},..,x_{j+l}) \parallel^2}{2\sigma_c^2}$}\right)}\, . 
\end{equation}

In other words, this kernel compares each substring $x_{i+1},x_{i+2},..,x_{i+l}$ of $\xb$ of size $l\leq L$ with each substring $x'_{j+1},x'_{j+2}. .., x'_{j+l}$ of $\xb'$ having the same length.
Each substring comparison yields a score that depends on the $\psib$-similarity of their respective amino acids and a shifting contribution term that decays exponentially rapidly with the distance between the starting positions of the two substrings.
The $ \sigma_p$ parameter controls the shifting contribution term.
The $\sigma_c$ parameter controls the amount of penalty incurred when the encoding vectors $\psib^l(x_{i+1}, .., x_{i+l})$ and $\psib^l(x'_{j+k},.., x'_{j+l})$ differ as measured by the squared Euclidean distance between these two vectors. 
The GS kernel outputs the sum of all the substring-comparison scores.

\begin{table}
\centering
 \begin{tabular}{ | l | l | l |}
  \hline
  Fixed parameters & Free parameters & Kernel name \\ \hline
  $ L=1, \sigma_p \to 0, \sigma_c \to 0$ &  & Hamming distance \\ \hline
  $ L \to \infty, \sigma_p \to 0, \sigma_c \to 0$ &  & Dirac delta \\ \hline
  $\sigma_p \to \infty, \sigma_c \to 0$ & $L$ & Blended Spectrum \cite{sc-04} \\ \hline
  $\sigma_p \to \infty$ & $L, \sigma_c$ & Blended Spectrum RBF \cite{twkr-10} \\ \hline
  $\sigma_c \to 0$ & $L, \sigma_p$ & Oligo \cite{Meinicke2004Oligo} \\ \hline
  $L \to \infty, \sigma_p \to 0$ &  $\sigma_c$ & Radial Basis Function (RBF) \\ \hline
  $\sigma_p \to 0, \sigma_c \to 0$ & $L$ & Weighted degree \cite{RS2004} \\ \hline
  $\sigma_p \to 0$ & $L, \sigma_c$ & Weighted degree RBF \cite{twkr-10} \\ \hline
  & $L, \sigma_p, \sigma_c$ & Generic String (GS) \\ \hline
 \end{tabular}
 \caption{Special cases of the GS kernel: eight know kernels can be obtained by fixing different parameters of the GS kernel.}
 \label{tbl:SpecialCases}
\end{table}

\bigskip
Also, note that the GS kernel can be used on strings of different lengths, which is a great advantage over a localized string kernel (of fixed length) such as the RBF and the weighted degree kernels~\cite{RS2004,twkr-10}.
In fact, the GS kernel generalizes eight known kernels. Table \ref{tbl:SpecialCases} lists them with the fixed and free parameters.
For example, when $\sigma_p$ approaches $+\infty$ and $\sigma_c$ approaches $0$, the GS kernel becomes identical to the blended spectrum kernel~\cite{sc-04}, which has a free parameter $L$ representing the maximum length of substrings. The free parameter values are usually determined by measuring the accuracy of the predictor on a separate (``hold-out'') part of the data that was not used for training, or by more elaborate sampling methods such as cross-validation.

\bigskip
In the next subsection, we prove that the GS kernel is symmetric positive semi-definite and, therefore, defines a scalar product in some large-dimensional feature space (see \cite{sc-04}). In other words, for any hyper-parameter values $(L, \sigma_p, \sigma_c)$, there exists a function $\phib_{\Xcal_{(L, \sigma_p, \sigma_c)}}$ transforming each finite sequence of amino acids into a vector such that $$GS(\xb,\xb', L, \sigma_p, \sigma_c)= \phib_{\Xcal_{(L, \sigma_p, \sigma_c)}}(\xb)\cdot\phib_{\Xcal_{(L, \sigma_p, \sigma_c)}}(\xb')\,.$$ 
Consequently, the solution minimizing the ridge regression functional $F(S,\wb)$ will be given by Equation~\eqref{eq:RRsolution} and is guaranteed to exist whenever the GS Kernel is used.

\subsubsection{Symmetric positive semi-definiteness of the GS kernel}
The fact that the GS kernel if positive semi-definite follows from the following theorem. The proof is provided as supplementary material [see Appendix]. 

\begin{theorem}\label{th:PSD}
Let $\Sigma$ be an alphabet (say the alphabet of all the amino acids).
For each $l\in\{1,..,L\}$, let $K_l:\Sigma^l\times\Sigma^l\longrightarrow \mathbb{R}$ be a symmetric positive semi-definite kernel. Let $A:\mathbb{R}\longrightarrow \mathbb{R}$ be any function that is the convolution of another function $B:\mathbb{R}\longrightarrow \mathbb{R}$, i.e., for all $z,z'\in\mathbb{R}$, we have
\begin{eqnarray*}
A(z-z')&=&\int_{-\infty}^{+\infty}\, B(z-t) B'(z'-t) \, \mathrm{d}t\,.
\end{eqnarray*}

\bigskip
\noindent
Then, the kernel $K$ defined, for any two strings on the alphabet $\Sigma$, as
\begin{eqnarray*}
K(\xb,\xb')&\eqdef& \sum_{l=1}^L\,\sum_{i=0}^{|\xb|-l}\,\sum_{j=0}^{|\xb'|-l} A(i-j) \ \, K_l((x_{i+1},..,x_{i+l})\,,\,(x'_{j+1},..,x'_{j+l}))
\end{eqnarray*}

is also symmetric positive semi-definite.

\end{theorem}

The positive semi-definiteness of the GS kernel comes from the fact that the GS kernel is a particular case of the more general kernel $K$ defined in the above theorem. Indeed, we first note that both kernels are identical except $A(i-j)$ in kernel $K$ is specialized to $\exp(\frac{-(i - j)^2}{2\sigma_p^2})$ in the GS kernel, and $K_l(\yb\,,\,\yb')$ in kernel $K$ is specialized to 
$\exp(\mbox{\large $\frac{-\parallel \psib^l(\yb) \,-\, \psib^l(\yb) \parallel^2}{2\sigma_c^2}$})$
in the GS kernel. Note that this last exponential is just an RBF kernel~(see \cite{sc-04} for a definition) that is defined over vectors of $\mathbb{R}^{ld}$ of the form $\psib^l(\yb)$; it is therefore positive semi-definite for any $l\in\{1,2,..,L\}$. For the first exponential, note that $\exp(\frac{-(i - j)^2}{2\sigma_p^2})$ is a function that is obtained from a convolution from another function since we can verify that  
$$
\exp\LP\frac{-(i - j)^2}{2\sigma_p^2}\RP\ =\  \frac{\sqrt{2}}{\sg_p\sqrt{\pi}}\int_{-\infty}^{+\infty} \exp\LP\frac{-(i - t)^2}{\sigma_p^2}\RP
\exp\LP\frac{-(j-t)^2}{\sigma_p^2}\RP dt\, .
$$ 
Indeed, this equality is a simple specialization of Equation~(4.13) of~\cite{rw-06}. It is related to the fact that the convolution of two Normal distributions is still a Normal distribution.

\bigskip
Finally, it is interesting to point out that Theorem~\ref{th:PSD} can be generalized to any function $A$ on measurable sets $M$ (not only the ones that are defined on $\mathbb{R}$), provided that $A$ is still is a convolution of another function $B:M\longrightarrow M$. We omit this generalized version in this paper since Theorem~\ref{th:PSD} suffices to prove that the GS kernel is positive semi-definite.

\subsubsection{Efficient computation of the GS kernel}
To cope with today's data deluge, the presented kernel should have a low computational cost. For this task, we first note that, before computing $GS(\xb,\xb', L, \sigma_p, \sigma_c) $ for each pair $(\xb,\xb')$ in the training set, we can first compute 
$$E(a,a')\quad \eqdef \quad \| \psib(a)-\psib(a') \|^2\quad = \quad \sum_{p=1}^d (\psi_p(a)-\psi_p(a'))^2 ,$$
for each pair $(a,a')$ of amino acids. After this pre-computation stage, done in $O(d\cdot|\Sigma|^2)$ time, each access to $E(a,a')$ is done in $O(1)$ time. We will not consider the running time of this pre-computation stage in the complexity analysis of the GS kernel, because it only has to be done once to be valid for any 5-tuple $(\xb,\xb',L,\sigma_p,\sigma_c)$. Let us recall that, in kernel methods, the kernel has to be evaluated a huge number of times, therefore any reasonable pre-computation can be omitted in the complexity analysis of such kernel.

Now, recall that we have defined $\psib^l:\Sigma^l\longrightarrow \mathbb{R}$ as the concatenation of vectors of the form $\psib(a)$ (see Equation~\eqref{eq:psib}). Hence  $\| \psib^l(\ab)-\psib^l(\ab') \|$ \, is an Euclidian norm, and we have that
\begin{equation}
\| \psib^l(\ab)-\psib^l(\ab') \|^2 \quad = \quad \sum_{k=1}^l \| \psib(a_k)-\psib(a_k') \|^2 \quad = \quad \sum_{k=1}^l E(a_k,a_k')
\end{equation}
Following this, we can now write the GS kernel as follows 
\begin{eqnarray}\label{eq:GS2}
GS(\xb,\xb', L, \sigma_p, \sigma_c) 
& = & 
\sum_{l=1}^{L}\sum_{i=0}^{|\xb|-l} \sum_{j=0}^{|\xb'|-l}\,  e^{\left(\mbox{\large $\frac{-(i - j)^2}{2\sigma_p^2}$}\right)} \ e^{\left(\mbox{\large $\frac{-\sum_{k=1}^{l}E(x_{i+k}, x'_{j+k})}{2\sigma_c^2}$}\right)}\\
\label{eq:GS3}
& = & 
\sum_{i=0}^{|\xb|} \sum_{j=0}^{|\xb'|}\,  e^{\left(\mbox{\large $\frac{-(i - j)^2}{2\sigma_p^2}$}\right)} \ \sum_{l=1}^{\min(L,|\xb|-i, |\xb'|-j)}\, 
e^{\left(\mbox{\large $\frac{-\sum_{k=1}^{l}E(x_{i+k}, x'_{j+k})}{2\sigma_c^2}$}\right)}\,,
\end{eqnarray}
where $\min(L,|\xb|-i, |\xb'|-j)$ is used in order to assure that ${i+k}$ and ${j+k}$ are valid positions in strings $\xb$ and $\xb'$. 

Now, for any $L, |\xb|, |\xb'|$, and any $i\in\{1,\ldots, |\xb|\}, j\in\{1,\ldots, |\xb'|\}$, let 
\begin{equation}\label{eq:matrixB}
 B_{i,j}\ \eqdef\ \sum_{l=1}^{\min(L,|\xb|-i, |\xb'|-j)}\, 
e^{\left(\mbox{\large $\frac{-\sum_{k=1}^{l}E(x_{i+k}, x'_{j+k})}{2\sigma_c^2}$}\right)}\, .
\end{equation}
We therefore have
\begin{equation}\label{eq:fastGS}
 GS(\xb,\xb', L, \sigma_p, \sigma_c) = \sum_{i=0}^{|\xb|} \sum_{j=0}^{|\xb'|} \exp\left(\frac{-(i - j)^2}{2\sigma_p^2}\right) \cdot B_{i,j}\, .
\end{equation}
Since $\min(L,|\xb|-i, |\xb'|-j)\leq L$, we see, from Equation~\eqref{eq:matrixB}, that the computation of each entry $B_{i,j}$ seems to involve $O(L^2)$ operations. However, we can reduce this complexity term to $O(L)$ by a dynamic programming approach.
Indeed, consider the following recurrence:
\begin{equation}
t_k = \begin{cases}
1 & \hbox{if } k=0\\
t_{k-1} \cdot e^{\left(\frac{-E(x_{i+k}, x'_{j+k})}{2\sigma_c^2} \right)} & \hbox{otherwise.} 
\end{cases}
\end{equation}
We thus have
\begin{equation}
B_{i,j} = \sum_{k=1}^{\min(L,|\xb|-i, |\xb'|-j)} t_k
\end{equation}
The computation of each entry $B_{i,j}$ therefore involves only $O(L)$ operations.
Consequently,  the running time complexity of  the GS kernel is  $O\left(|\xb| \cdot |\xb'| \cdot L\right)$.

\subsubsection{GS Kernel relaxation}
We now show how to compute a very close approximation of the GS kernel in linear time.
Such a feature is interesting if one wishes to do a pre or post treatment where the symmetric positive semi-definite (SPSD) property of the kernel is not required.
For example, as opposed to the training stage where the inverse of $\Kb + \Ib/C$ is guaranteed to exists only for a SPSD matrix $\Kb$, kernel values in the prediction stage could be approximate.
Indeed, the scalar product with $\alb$ is defined for non positive semi-definite kernel values.
This scheme would greatly speed up the predictions with a very small precision lost.

The shifting penalizing term $\exp\left(\frac{-(i - j)^2}{2\sigma_p^2}\right)$ in Equation (\ref{GSkernel}) implies that the further two substrings are from each other, no matter how similar they are, their contribution to the kernel will vanish exponentially rapidly.
Let $\delta$ be the maximum distance between two substrings that we intend to consider in the computation of the approximate version of the GS kernel. In other words, any substring whose distance is greater than $\delta$ will not contribute.
We propose to fix $\delta = \lceil3\sigma_p\rceil$. In this case, the contribution of any substring beyond $\delta$ is bound to be minimal.
For the purpose of demonstration, let $P$ be the  $|\xb| \times |\xb'|$ matrix 
\begin{equation}
  P_{i,j} \eqdef \left\{ 
  \begin{array}{l l}
    0 & \quad \text{if $| i-j | > \delta$}\\
    \exp\left(\frac{-(i - j)^2}{2\sigma_p^2}\right) & \quad \text{otherwise}\, .\\
  \end{array} \right.
\end{equation}
$P$ is thus a sparse matrix with exactly $\delta |\xb| + \delta |\xb'| - \delta^2$ non-zero values around it's diagonal.
We can therefore write this approximation of the GS kernel as
\begin{equation}
 GS'(\xb,\xb', L, \sigma_p, \sigma_c, \delta) = \sum_{i=0}^{|\xb|} \sum_{j=0}^{|\xb'|} P_{i,j} \cdot B_{i,j}\, .
\end{equation} 
It is now clear that only values of $B$ for which the value in $P$ is non-zero need to be computed.
The complexity of $GS'$ is dominated by the computation of matrix $B$ whose $\delta |\xb| + \delta |\xb'| - \delta^2$ entries can be computed in $O\left(\max(|\xb|, |\xb'|)\right)$.
Since $L$ and $\delta$ are constant factors, we have that $GS' \in O\left(\max(|\xb|, |\xb'|)\right)$ with an  optimal linear complexity. 

\subsection{Kernel for protein binding pocket}\label{sec:binding}

Hoffmann et al. \cite{Hoffmann2010New} proposed a new similarity measure between protein binding pockets. The similarity measure aligns atoms extracted from the binding pocket in $3D$ and assigns a score to the alignment. Pocket alignment is possible for proteins that share low sequence and structure similarity. They proposed two variations of the similarity measure. The first variation only compares the shape of pockets to assign a score. In the second variation, atom properties, such as partial charges, re-weight the contribution of each atom to the score. We will refer to these two variations respectively as sup-CK and sup-CK$_L$. Since both scores are invariant by rotation and translation, they are not positive semi-definite kernels. To obtain a valid kernel, we have used the so-called empirical kernel map where each $\yb$ is mapped explicitly to $(k(\yb_1, \yb), k(\yb_2, \yb),\ldots,k(\yb_m, \yb))$. To ensure reproducibility and avoid implementation errors, all experiments were done using the implementation provided by the authors. An illustration of the pocket creation for the SupCk kernel is shown in Figure~\ref{fig:bindingPocket}.

\begin{figure}
\begin{center}
\includegraphics[scale=0.4]{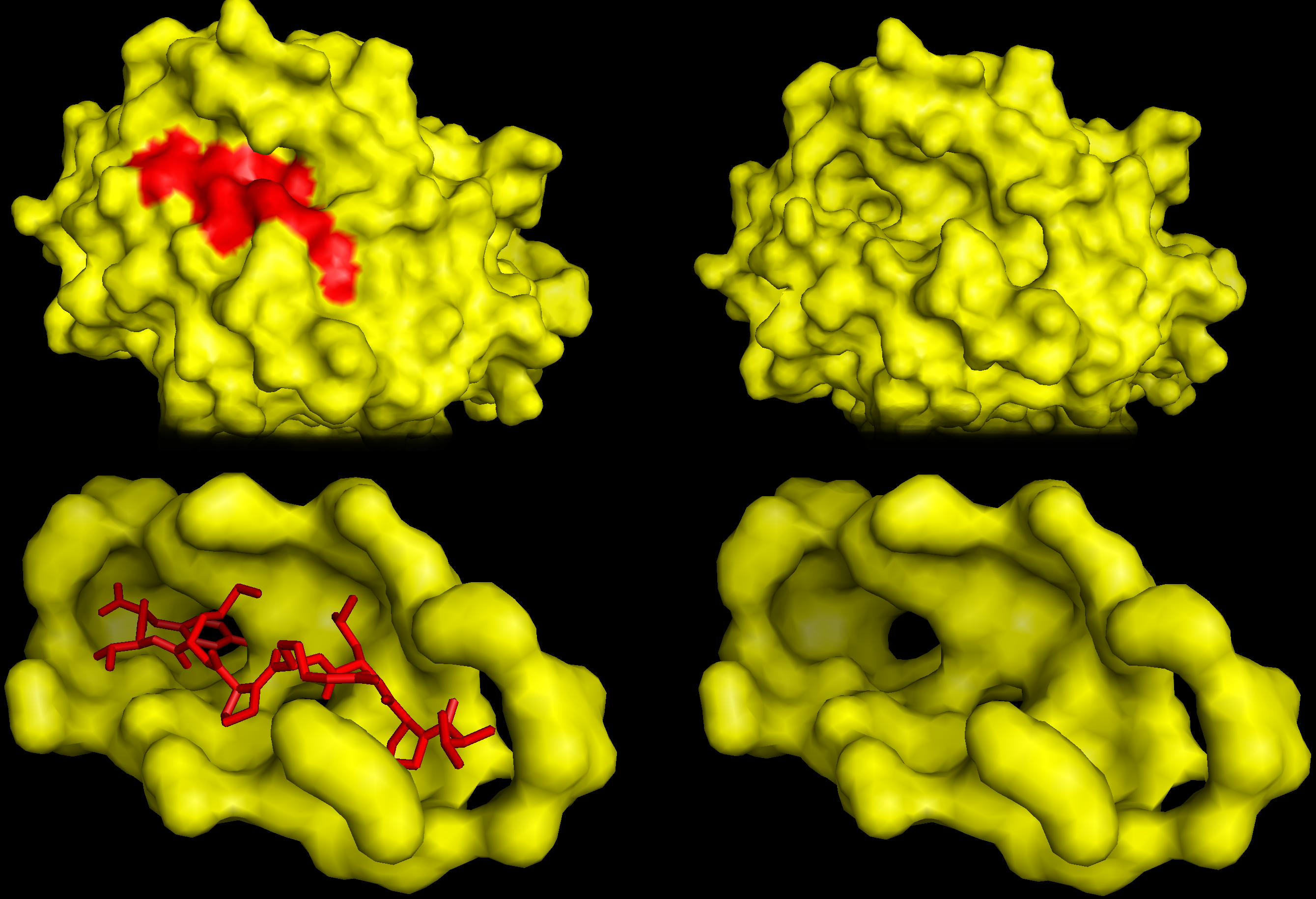}
\caption{A pyMOL illustration of the binding pocket kernel shows a MHC-I molecule B*3501 complexed with a peptide (VPLRPMTY) from the NEF protein of HIV1 (PDB ID 1A1N). The MHC protein is shown in yellow, the peptide in shown in red.}
\label{fig:bindingPocket}
\end{center}
\end{figure}

\subsection{Metrics and experimental design}\label{sec:Framework}

When dealing with regression values, classical metrics used for classification such as the area under the ROC curve (AUC)~\cite{Swets1988} are not suitable.
To compute the AUC, some authors fix a binding threshold value used to transform the regression problem into a binary classification problem.
The real value output of the predictor is also mapped to a binary class using the same threshold value. The AUC is then computed on the binary output.
Unfortunately, this method makes the AUC metric dependent on the threshold value chosen to transform the real output values in the regression problem. For this reason, we decided not to present AUC results in the main paper. Those results are nevertheless provided as supplementary material [see Appendix]. 

Metrics such as the root mean squared error (RMSE), the coefficient of determination ($R^2$) and the Pearson product-moment correlation coefficient (PCC) are more suited for regression. In this paper, we have used the PCC and the RMSE to evaluate the performance of our method.

Except when otherwise stated, nested cross-validation was done for estimating the PCC and the RMSE of the predicted binding affinities. For all $n$ outer folds, $n-1$ inner cross-validation folds were used for the selection of the kernel hyper-parameters and the $C$ parameter of Equation~\eqref{eq:RRsolution}. All reported values are computed on the union of the outer fold test set predictions. This is important since an average of correlation coefficients is not a valid correlation coefficient. This is also true for the root mean squared error. An illustration of the nested cross-validation procedure is shown in Figure~\ref{fig:nestedCV}.

More precisely, let $\bar e$ denote the average affinity in the data set $\Dcal$. Let $T_k$ for $k\in\{1,\ldots,10\}$ denote the testing set of the $k^{\mbox{\footnotesize th}}$ outer fold and let $h_{\Dcal\setminus T_k}(\xb_i, \yb_i)$ be the predicted binding affinity on example $((\xb_i,\yb_i), e_i)$ of the predictor built from $\Dcal\setminus T_k$. Then the correlation coefficient was computed using

\begin{equation}
 PCC =  \sqrt{ \frac{ \sum_{k=1}^{n} \sum_{i\in T_k} \LP e_i - h_{\Dcal\setminus T_k}(\xb_i,\yb_i)\RP^2 }{ \sum_{i\in\Dcal}\LP e_i - \bar e\RP^2 }}\, .
\end{equation}

An algorithm that, on average, makes the same error as $\bar e$ will give $PCC = 0$ and an algorithm that always returns a perfect predictor will give $PCC = 1$.

As for the RMSE, it was computed using

\begin{equation}
RMSE = \sqrt{\frac{\sum_{i\in T_k} (e_i - h_{\Dcal\setminus T_k}(\xb_i,\yb_i))^2}{|T_k|}}
\end{equation}

Therefore, the perfect predictor will give $RMSE=0$ and the value of this metric will increase as the quality of the predictor decrease.

All the p-values reported in this article were computed using the two-tailed Wilcoxon signed-ranked test.

\begin{figure}
\begin{center}
\includegraphics[scale=0.4]{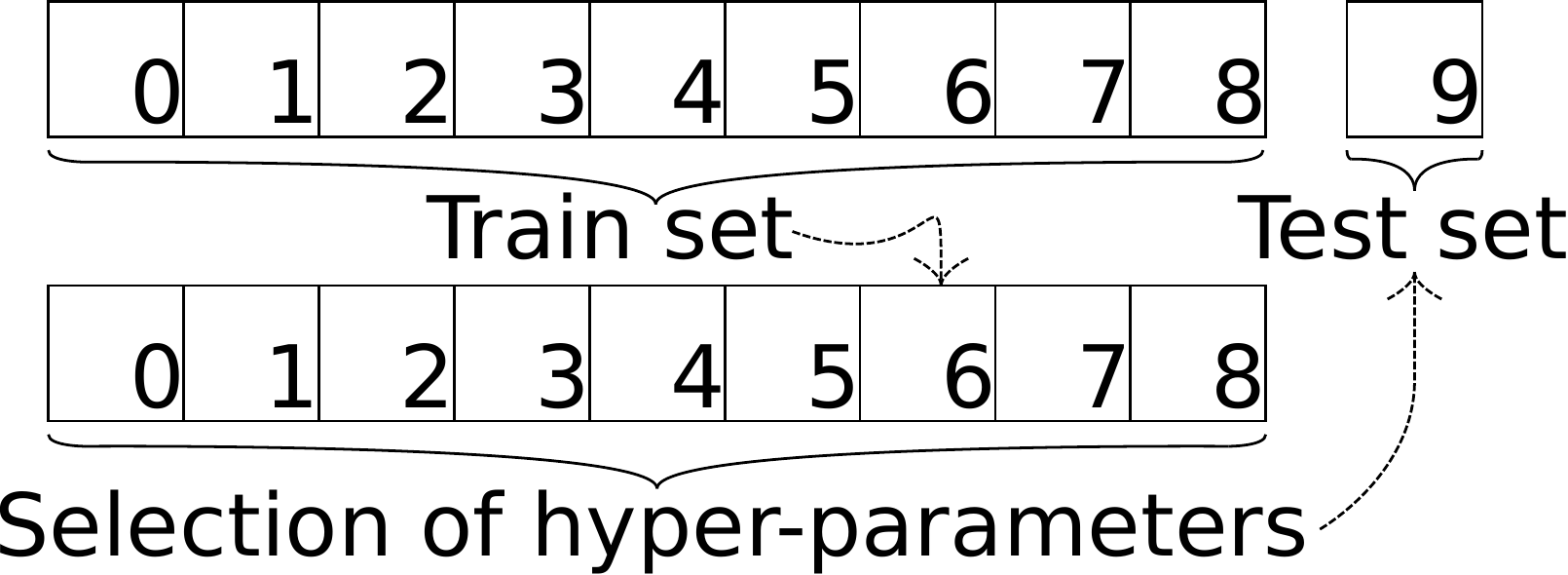}
\end{center}
\caption{Illustration of the nested $10$-fold cross-validation procedure. For all of the $10$ outer folds, an inner $9$ fold cross-validation scheme was used to select hyper-parameters.}
\label{fig:nestedCV}
\end{figure}

\subsection{Data}
\subsubsection{PepX database}

The PepX database \cite{Vanhee2010PepX} contains $1431$ high-quality peptide-protein complexes along with their protein and peptide sequences, high quality crystal structures, and binding energies (expressed in kcal/mol) computed using the FoldX force field method.
Full diversity of structural information on protein-peptide complexes is achieved with peptides bound to, among others, MHC, thrombins, $\alpha$-ligand binding domains, SH3 domains and PDZ domains.
This database recently drew attention in a review on the computational design of peptide ligands \cite{Vanhee2011Computational} where it was part of large structural studies to understand the specifics of peptide binding.
A subset of $505$ non-redundant complexes was selected based on the dissimilarity of their binding interfaces. The authors of the database done the selection in such a way that
this smaller subset still represented the full diversity of structural information on peptide-protein complexes present in the entire Protein Data Bank (PDB), see \cite{Vanhee2010PepX} for a description of the method.
We will refer to the smaller subset as the ``PepX Unique'' data set and to the whole data base as ``PepX All''.

The few complexes with positive binding energies were removed from the dataset.
No other modifications were made on the original database.

\subsubsection{Major histocompatibility complex class II (MHC-II)}\label{seq:MHCIIResults}

Two different approaches were used for the prediction of MHC class II - peptide binding affinities: single-target and multi-target (pan-specific).

Single-target prediction experiments were conducted using the data from the IEDB dataset proposed by the authors of the RTA method \cite{Bordner2010RTA}. The latter consists of 16 separate datasets, each containing data on the peptides binding to an MHC class II allotype. For each allotype, the corresponding dataset contains the binding peptide sequences and their binding affinity in kcal/mol. These datasets have previously been separated into 5 cross-validation folds to minimize overlapping between peptide sequences in each fold. It is well known  in the machine learning community that such practice should be avoided, as opposed to random fold selection, since the training and test sets are then not independently generated. These predefined folds were nevertheless used for the purpose of comparison with other learning methodologies that have used them.

Pan-specific experiments were conducted on the IEDB dataset proposed by the authors of the NetMHCIIpan method \cite{Nielsen2008}. The dataset contains 14 different HLA-DR allotypes, with $483$ to $5648$ binding peptides per allotype. For each complex, the dataset contains the HLA allele's identifier (e.g.: \textit{DRB1*0101}), the peptide's sequence and the $\log 50k$ transformed IC50 (Inhibitory Concentration $50\%$), which is given by $1-\log_{50000} IC50$.

As pan-specific learning requires comparing HLA alleles using a kernel, the allele identifiers contained in the dataset were not directly usable for this purpose. 
Hence, to obtain a useful similarity measure (or kernel) for pairs of HLA alleles, we have used the pseudo sequences composed of the amino acids at highly polymorphic positions in the alleles' sequences. These amino acids are potentially in contact with peptide binders\cite{Nielsen2008}, therefore contributing to the MHC molecule's binding specificity.
The authors of the NetMHCIIpan method proposed using pseudo sequences composed of the amino acids at 21 positions that were observed to be polymorphic for HLA-DR, DP and DQ \cite{Nielsen2008}. With respect to the IMGT nomenclature \cite{Robinson2000}, these amino acids are located between positions 1 and 89 of the MHC's $\beta$ chain.
Pseudo sequences consisting of all 89 amino acids between these positions were also used to conduct the experiments.

\subsubsection{Quantitative structure affinity model (QSAM) benchmark}

Three well-studied benchmark datasets for designing quantitative structure affinity models were also used to compare our approach: 58 angiotensin-I converting enzyme (ACE) inhibitory dipeptides, 31 bradykinin-potentiating pentapeptides and 101 cationic antimicrobial pentadecapeptides.
These data sets were recently the subject of extensive studies \cite{Zhou2010Gaussian} where partial least squares (PLS), Artificial Neural Networks (ANN), Support Vector Regression (SVR), and Gaussian Processes (GP) were used to predict the biological activity of the peptides.
GP and SVR were found to have the best results on the testing set, but their experiment protocol was unconventional because the training and test sets were not randomly selected from the data set. Instead,  their testing examples were selected from a cluster analysis performed on the whole data set---thus favoring learning algorithms that tend to cluster their predictions according to the same criteria used to split the data.
Instead, we randomly selected the testing examples from the whole data set---thus favoring no algorithm \emph{a priori}. 
Theses datasets were chosen to demonstrate the ability of our method to learn on both small and large datasets.

\section{Results and discussion}

\subsection{PepX database}

To our knowledge, this is the first kernel method attempt at learning a predictor which takes the protein crystal and the peptide sequence as input to predict the binding energy of the complex.
Many consider this task as a major challenge with important consequences across biology.
String kernels such as the LA-kernel and the blended spectrum were used while conducting experiments on proteins.
They did not yield good results (data not shown), mainly because they do not consider the protein's secondary structure information.
To validate this hypothesis and improve our results, we tried using the MAMMOTH kernel, a similarity function proposed by \cite{Ortiz2002}, recently used for prediction of protein-protein interactions from structures \cite{Hue2010Largescale}.
The MAMMOTH uses structural information from crystals to aligns two proteins using their secondary structure and assigns a score to the alignment.
The greater the similarity between two proteins' secondary structure, the greater the alignment score will be.
The use of the MAMMOTH kernel did improve the results but was still missing an important aspect of protein-peptide interaction.
The interaction takes place at a very specific location on the surface of the protein called the binding pocket.
Two proteins may be very different, but if they share a common binding pocket, it is likely that they will bind similar ligands.
This is the core idea that motivated \cite{Hoffmann2010New} in the creation of the sup-CK binding pocket kernel.

Choosing a kernel for the peptides was also a challenging task.
Sophisticated kernels for local signals such as the RBF, weighted degree, and weighted degree RBF could not be used because peptide lengths were not equal. In fact, peptide lengths vary between 5 and 35 amino acids, which makes the task of learning a predictor and designing a kernel even more challenging.
This was part of our motivation in designing the GS kernel.
For all experiments, the BLOSUM $50$ matrix was found to be the optimal amino acid descriptors during cross-validation.

\begin{table}
\center
  \begin{tabular}{l|c|c|c|c|c|}
  \multicolumn{6}{}{} \\
  \cline{2-6}
  \multirow{3}{*}{} & \multicolumn{1}{c|}{SVR} & \multicolumn{4}{c|}{KRR} \\
  \cline{2-6}
  & sup-CK & \multicolumn{2}{c|}{sup-CK} & \multicolumn{2}{c|}{sup-CK$_L$} \\
  \cline{2-6}
  & BS & BS & GS &  BS & GS \\ \hline
  \multicolumn{0}{|l|}{PepX Unique} & $0.6822$ & $0.7072$ & $\textbf{0.7300}$ & $0.7110$ & $0.7264$ \\ \hline
  \multicolumn{0}{|l|}{PepX All}    & $0.8227$ & $0.8580$ & $0.8648$ 		  & $0.8601$ & $\textbf{0.8652}$ \\ \hline
  \end{tabular}
  \caption{Correlation coefficient (PCC) for multiple target predictions on the PepX database. Best results are highlighted in bold.}
  \center
  \label{tbl:resultsMultipleTarget}
\end{table}

Table \ref{tbl:resultsMultipleTarget} presents the first machine learning results for the prediction of binding affinity given any peptide-protein pair.
We first observe that KRR has better accuracy than SVR. 
We also note that using the GS kernel over the simpler BS kernel improve accuracy for both the sup-CK and the sup-CK$_L$ kernels for binding pockets.
It is surprising that the sup-CK$_L$ kernel does not outperform the sup-CK kernel on both benchmarks, since the addition of the atom partial charges should provide more relevant information to the predictor. 

\begin{figure}
\begin{center}
\subfigure[PepX Unique]{
\label{fig:pepxUnique_illustration}
\includegraphics[scale=0.4]{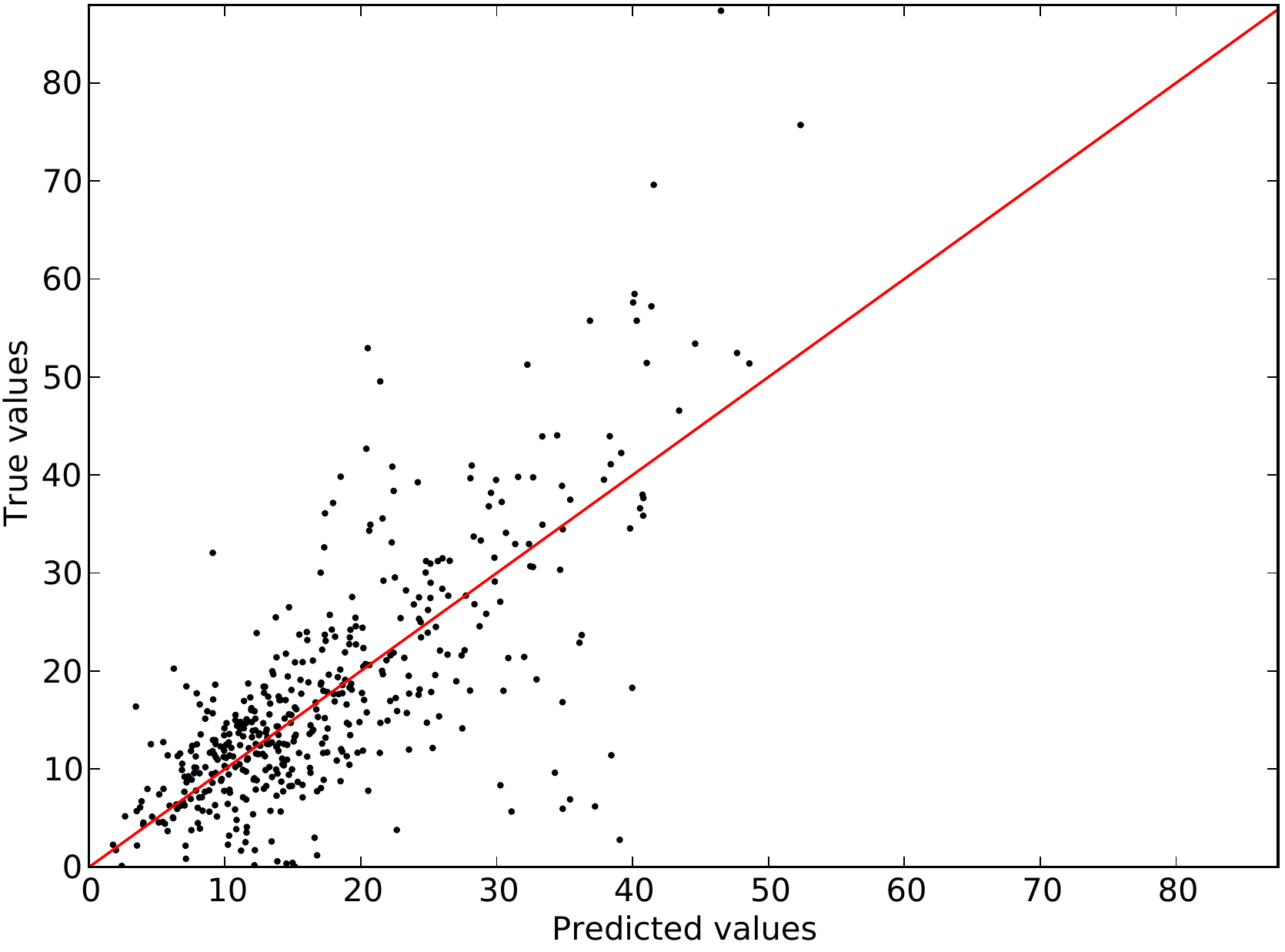}
}
\subfigure[PepX All]{\label{fig:pepxAll_illustration}
\includegraphics[scale=0.4]{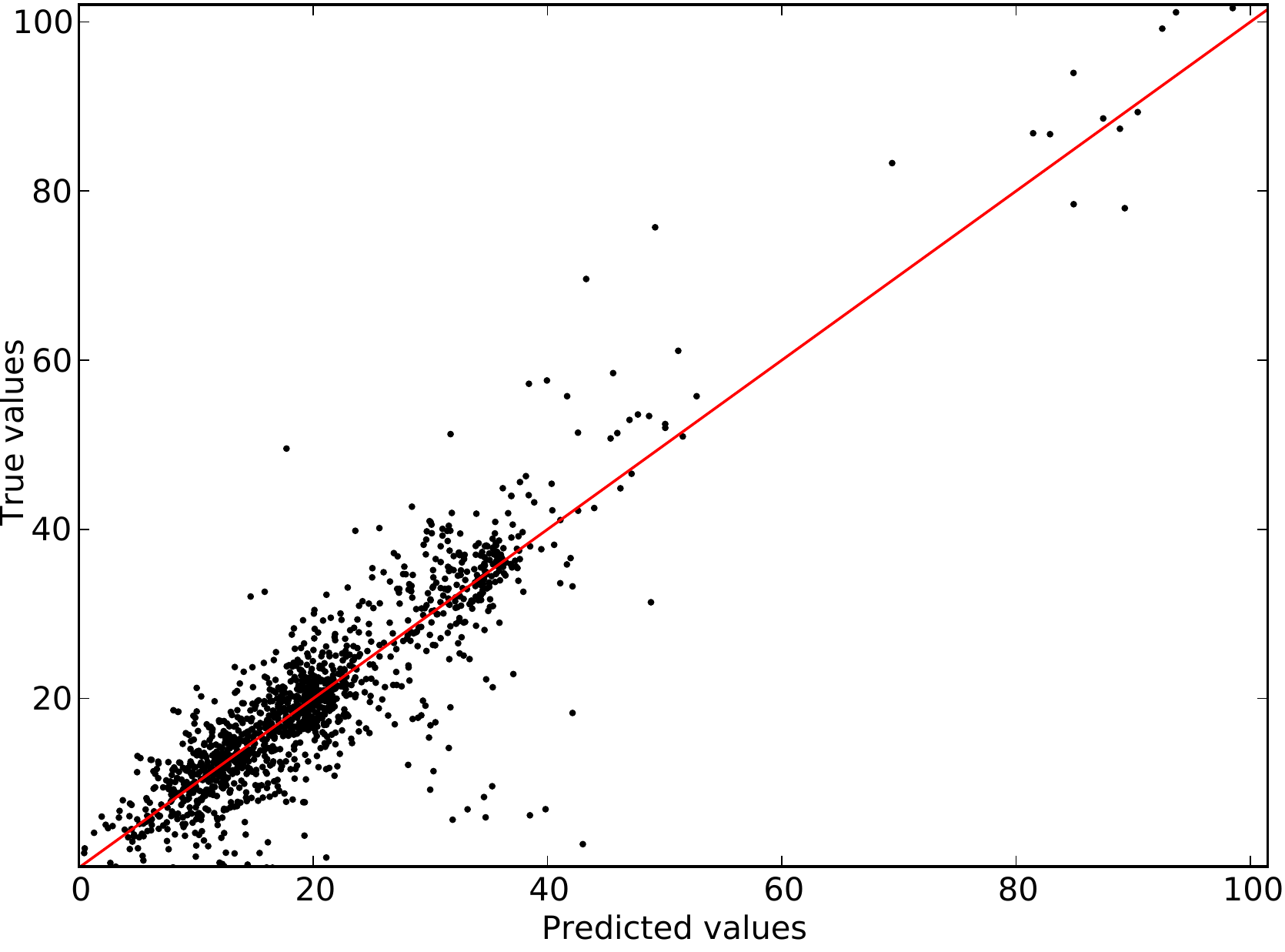}
}
\end{center}
\caption{Predicted values as a function of the true values for the PepX Unique (left) and PepX All (right) datasets. A perfect predictor would have all it's predictions lying on the $x=y$ red line.}
\end{figure}

Figures \ref{fig:pepxUnique_illustration} and \ref{fig:pepxAll_illustration} present an illustration of the prediction accuracy using sup-CK for the PepX Unique dataset and sup-CK$_L$ for the PepX All dataset.
For illustration purposes, the absolute value of the binding energy has been plotted.
We observe that the predictor has the property of maintaining the ranking, such that the peptides with high binding affinity can be identified, an important feature for drug discovery.
Peptides with the highest affinities are the ones that, ultimately, will serve as precursor drug or scaffold in a rational drug design program.

Experiments showed that a Pearson correlation coefficient of $\approx 1.0$ is attainable on the training set when using the binding pocket kernel, the GS kernel and a large $C$ value (empirically $\approx 100$).
This is a strong indication that our method has the ability of building a perfect predictor, but the lack of data quality and quantity may be responsible for the reduced performance on the testing set.
Hence better data will improve the quality of the predictor.
Initially, biological validation will be necessary but ultimately, when sufficient data is gathered, the predictor will provide accurate results that are currently only achievable by high cost biological experimentation.

\subsection{Major histocompatibility complex class II (MHC-II)}

\subsubsection{Single-target predictions}

We performed a single-target prediction experiment using the dataset proposed by the authors of the RTA method \cite{Bordner2010RTA}.
The goal of such experiments was to evaluate the ability of a predictor to predict the binding energy (kcal/mol) of an unknown peptide to a specific MHC allotype when training only on peptides binding to this allotype.
For each of the 16 MHC allotypes, a predictor was trained using kernel ridge regression with the GS kernel and a nested cross-validation scheme was used.
For comparison purposes, the nested cross-validation was done using the 5 predefined cross-validation folds provided in \cite{Bordner2010RTA}.
Again, this is sub-optimal from the statistical machine learning perspective, since the known guarantees on the risk of a predictor~\cite{sc-04,ss-02} normally require that the examples be generated independently from the same distribution. 

Three common metrics were used to compare the methods: the Pearson correlation coefficient (PCC), the root mean squared error (RMSE), and the area under the ROC curve (AUC).
The PCC and the RMSE results are presented in Table \ref{tbl:resultsHLASingleTargetSummary}, AUC values can be found as supplementary material [see Appendix].
The PCC results show that our method significantly outperforms the RTA method on $13$ out of $16$ allotypes with a p-value of $0.0308$.
The inferior results for certain allotypes may be attributed to the small size of these datasets.
In addition, the RMSE results show that our method clearly outperforms the RTA method on all $16$ allotypes with a p-value of $0.0005$.\\

\begin{table}
	\center
	\begin{tabular}{l|cc|cc|c}
	\multicolumn{1}{c}{} & \multicolumn{2}{c}{\textbf{PCC}} & \multicolumn{2}{c}{\textbf{RMSE (kcal/mol)}} & \multicolumn{1}{c}{}\\
	\cmidrule(r){2-3}\cmidrule(r){4-5}
	\multicolumn{1}{l}{\textbf{MHC $\boldsymbol{\beta}$ chain}} & \multicolumn{1}{c}{\textbf{KRR+GS}}  & \multicolumn{1}{c}{\textbf{RTA}} & \multicolumn{1}{c}{\textbf{KRR+GS}} & \multicolumn{1}{c}{\textbf{RTA}} & \multicolumn{1}{c}{\textbf{\# of examples}}\\ 
	\hline
	DRB1*0101   & \textbf{0.632} & 0.530 & \textbf{1.20} & 1.43 & 5648\\
	DRB1*0301   & \textbf{0.538} & 0.425 & \textbf{1.16} & 1.46 & 837\\
	DRB1*0401   & \textbf{0.430} & 0.340 & \textbf{1.44} & 1.72 & 1014\\
	DRB1*0404   & \textbf{0.491} & 0.487 & \textbf{1.25} & 1.38 & 617\\
	DRB1*0405   & \textbf{0.530} & 0.442 & \textbf{1.09} & 1.35 & 642\\
	DRB1*0701   & \textbf{0.645} & 0.484 & \textbf{1.24} & 1.62 & 833\\
	DRB1*0802   & \textbf{0.469} & 0.412 & \textbf{1.19} & 1.34 & 557\\
	DRB1*0901   & 0.303 & \textbf{0.369} & \textbf{1.55} & 1.68 & 551\\
	DRB1*1101   & \textbf{0.550} & 0.450 & \textbf{1.17} & 1.45 & 812\\
	DRB1*1302   & \textbf{0.468} & 0.464 & \textbf{1.51} & 1.64 & 636\\
	DRB1*1501   & \textbf{0.502} & 0.438 & \textbf{1.41} & 1.53 & 879\\
	DRB3*0101   & 0.380 & \textbf{0.425} & \textbf{1.03} & 1.13 & 483\\
	DRB4*0101   & \textbf{0.613} & 0.522 & \textbf{1.10} & 1.33 & 664\\
	DRB5*0101   & \textbf{0.541} & 0.434 & \textbf{1.20} & 1.57 & 835\\
	H2*IA$_b$   & \textbf{0.603} & 0.556 & \textbf{1.00} & 1.15 & 526\\
	H2*IA$_d$   & 0.325 & \textbf{0.563} & \textbf{1.44} & 1.53 & 306\\
	\hline 
	\multicolumn{1}{r|}{Average:}    & \textbf{0.501} & 0.459 & \textbf{1.25} & 1.46 & \\
	\hline
	\end{tabular}
	\caption{Comparison of HLA-DR prediction results on the dataset proposed by the authors of RTA. Best results for each metric are highlighted in bold.}
	\label{tbl:resultsHLASingleTargetSummary}
\end{table}

\subsubsection{Pan-specific predictions}

To further evaluate the performance of our method and the potential of the GS kernel, pan-specific predictions were performed using the dataset proposed by the authors of NetMHCIIpan \cite{Nielsen2008}.
The authors proposed a new cross-validation scheme called the \emph{leave one allele out} (LOAO) where all but one allele are used as training set and the remaining allele is used as testing set.
This is a more challenging problem, as the predictor needs to determine the binding affinity of peptides for an allele which was absent in the training data. 
The binding specificity of an allele's interface is commonly characterized using a pseudo sequence extracted from the beta chain's sequence \cite{Zhang2011Toward, Nielsen2008, Nielsen2010}.
During our experiments the 21 amino acid pseudo sequences were found to be optimal.
The 89 amino acid pseudo sequences yielded similar, but slightly suboptimal results.
For all experiments, the GS kernel was used for the allele pseudo sequences and for the peptide sequences.
All results were obtained with the same LOAO scheme presented in \cite{Nielsen2008}. 
For each allele an inner LOAO cross-validation was done for the selection of hyper-parameters.

To assess the performance of the proposed method, the PCC and the RMSE results are shown in Table \ref{tbl:resultsHLAMultiTargetSummary}, AUC values can be found in the supplementary material [see Appendix]. 
Since we performed LOAO cross-validation, the PCC, RMSE and AUC values were calculated on each test fold individually, thus yielding results for each allele.

The PCC results show that our method outperforms the MultiRTA\cite{Bordner2010MultiRTA} (p-value of $0.001$) and the NetMHCIIpan-2.0\cite{Nielsen2010} (p-value of $0.0574$) methods. 
Since the dataset contained values in $\log 50k$ transformed IC50 (Inhibitory Concentration $50\%$), the calculation of the RMSE values required converting the predicted values to kcal/mol using the method proposed in \cite{Bordner2010RTA}.

The RMSE values are only shown for our method and the MultiRTA method, since such values were not provided by the authors of NetMHCIIpan-2.0. The RMSE results indicate that our method globally outperforms MultiRTA with a p-value of $0.0466$.

\begin{table}
	\center
	\tabcolsep=0.09cm
	\begin{tabular}{l|ccc|cc|c}
	\multicolumn{1}{c}{} & \multicolumn{3}{c}{\textbf{PCC}} & \multicolumn{2}{c}{\textbf{RMSE (kcal/mol)}} & \multicolumn{1}{c}{}\\
	\cmidrule(r){2-4}\cmidrule(r){5-6}
	\multicolumn{1}{l}{\textbf{MHC $\boldsymbol{\beta}$ chain}} & \multicolumn{1}{c}{\textbf{KRR+GS}}  & \multicolumn{1}{c}{\textbf{MultiRTA}} & \multicolumn{1}{c}{\textbf{NetMHCIIpan-2.0}} & \multicolumn{1}{c}{\textbf{KRR+GS}} & \multicolumn{1}{c}{\textbf{MultiRTA}} & \multicolumn{1}{c}{\textbf{$\#$ of examples}}\\ 
	\hline
	DRB1*0101 & \textbf{0.662} & 0.619 & 0.627 & 1.48 & \textbf{1.33} & 5166\\
	DRB1*0301 & \textbf{0.743} & 0.438 & 0.560 & \textbf{1.29} & 1.36 & 1020\\
	DRB1*0401 & \textbf{0.667} & 0.534 & 0.652 & \textbf{1.36} & 1.56 & 1024\\
	DRB1*0404 & 0.709 & 0.623 & \textbf{0.731} & \textbf{1.18} & 1.33 & 663\\
	DRB1*0405 & 0.606 & 0.566 & \textbf{0.626} & \textbf{1.25} & 1.28 & 630\\
	DRB1*0701 & 0.694 & 0.620 & \textbf{0.753} & \textbf{1.34} & 1.51 & 853\\
	DRB1*0802 & \textbf{0.728} & 0.523 & 0.700 & \textbf{1.23} & 1.45 & 420\\
	DRB1*0901 & 0.471 & 0.375 & \textbf{0.474} & \textbf{1.53} & 2.01 & 530\\
	DRB1*1101 & \textbf{0.786} & 0.603 & 0.721 & \textbf{1.16} & 1.46 & 950\\
	DRB1*1302 & \textbf{0.416} & 0.365 & 0.337 & 1.73 & \textbf{1.68} & 498\\
	DRB1*1501 & \textbf{0.612} & 0.513 & 0.598 & \textbf{1.46} & 1.57 & 934\\
	DRB3*0101 & \textbf{0.654} & 0.603 & 0.474 & 1.52 & \textbf{1.10} & 549\\
	DRB4*0101 & \textbf{0.540} & 0.508 & 0.515 & \textbf{1.41} & 1.61 & 446\\
	DRB5*0101 & \textbf{0.732} & 0.543 & 0.722 & \textbf{1.28} & 1.60 & 924\\
	\hline 
	\multicolumn{1}{r|}{Average:} & \textbf{0.644} & 0.531 & 0.606 & \textbf{1.37} & 1.49 & \\
	\hline
	\end{tabular}
	\caption{Comparison of pan-specific HLA-DR prediction results on the dataset proposed by the authors of NetMHCIIpan. Best results for each metric are highlighted in bold.}
	\label{tbl:resultsHLAMultiTargetSummary}
\end{table}

\subsection{Quantitative structure affinity model (QSAM) benchmark}

For all datasets, the extended $z$ scale \cite{Zhou2010Gaussian} was found to be the optimal amino acids descriptors during cross-validation. All the results in this section were thus obtained using the extended $z$ scale for the RBF and GS kernels.
All peptides within each data set are of the same length, which is why the RBF kernel can be applied, as opposed to the PepX database or the two MHC-II benchmark datasets.
Note the RBF kernel is a special case of the GS kernel. Hence, the results obtained from our method using the GS kernel were likely to be at least as good as those obtained with the RBF kernel.

Table \ref{tbl:resultsSingleTarget} present the results obtained when applying the method from \cite{Zhou2010Gaussian} (SVR learning with the RBF kernel) and our method (KRR learning with the GS kernel). Results with the RBF kernel and KRR are also presented to illustrate the gain in accuracy obtained from the more general GS kernel.

We observed that kernel ridge regression (KRR) had a slight accuracy advantage over support vector regression (SVR). Moreover, SVR has one more hyperparameter to tune than KRR: the $\epsilon$-insensitive parameter. Consequently, KRR should be preferred over SVR for requiring a substantially shorter learning time.
Also, we show in Table \ref{tbl:resultsSingleTarget} that the GS kernel outperforms the RBF kernel on all three QSAM data sets (when limiting ourself to KRR). Considering these results, KRR with the GS kernel clearly outperforms the method of \cite{Zhou2010Gaussian} on all data sets.

\begin{table}
\center
   \begin{tabular}{l|c|c|c|}
    \cline{2-4}
    \multirow{2}{*}{} & \multicolumn{1}{c|}{SVR} & \multicolumn{2}{c|}{KRR} \\
    \cline{2-4}
    & RBF & RBF & GS  \\ \hline
    \multicolumn{0}{|l|}{ACE}           & $0.8782$ & $0.8807$ & $\textbf{0.9044}$  \\ \hline
    \multicolumn{0}{|l|}{Bradykinin}    & $0.7491$ & $0.7531$ & $\textbf{0.7641}$  \\ \hline
    \multicolumn{0}{|l|}{Cationic}      & $0.7511$ & $0.7417$ & $\textbf{0.7547}$  \\ \hline
  \end{tabular}
  \caption{Correlation coefficient (PCC) on the QSAM benchmarks. Best results are highlighted in bold.}
  \label{tbl:resultsSingleTarget}
\end{table}

\section{Conclusions}

We have proposed a new kernel designed for small bio-molecules (such as peptides) and pseudo-sequences of binding interfaces.
The GS kernel is an elegant generalization of eight known kernels for local signals.
Despite the richness of this new kernel, we have provided a simple and efficient dynamic programming algorithm for its exact computation and a linear time algorithm for its approximation.
Combined with the kernel ridge regression learning algorithm and the binding pocket kernel, the proposed kernel yields biologically relevant results on the PepX database.
For the first time, a predictor capable of accurately predicting the binding affinity of any peptide to any protein was learned using this database.
Our method significantly outperformed RTA on the single-target prediction of MHC-II binding peptides.
Impressive state-of-the-art results were also obtained on the pan-specific MHC-II task, outperforming both MultiRTA and NetMHCIIpan-2.0.
Moreover, the method was successfully tested on three well studied datasets for the quantitative structure affinity model.

A predictor trained on the whole IEDB database or PDB database, as opposed to benchmark datasets, would be a substantial tool for the community.
Unfortunately, learning a predictor on very large datasets (over $25000$ examples) is still a major challenge with most machine learning methods, as the similarity (Gram) matrix becomes hard to fit into the memory of most computers.
We propose to expand the presented method to very large datasets as future work.

\section*{Author's contributions}
SG designed the GS kernel, algorithms for it's computation, implemented the learning algorithm and conducted experiments on the PepX and QSAM datasets.
MM designed the learning algorithm.
FL and MM did the proof of the symmetric positive semi-definiteness of the GS kernel.
AD conducted experiments on MHC-II datasets.
JC provided biological insight and knowledge.
This work was done under the supervision of MM, FL and JC.
All authors contributed to, read and approved the final manuscript.

\section*{Acknowledgements}
  \ifthenelse{\boolean{publ}}{\small}{}
Computations were performed on the SciNet supercomputer at the University of Toronto, under the auspice of Compute Canada. The operations of SciNet are funded by the Canada Foundation for Innovation (CFI), the Natural Sciences and Engineering Research Council of Canada (NSERC), the Government of Ontario and the University of Toronto.
JC is the Canada Research Chair in Medical Genomics. This work was supported in part by
the Fonds de recherche du Québec - Nature et technologies (FL, MM \& JC; 2013-PR-166708) and the NSERC Discovery Grants (FL; 262067, MM; 122405).
 
\newpage
{\ifthenelse{\boolean{publ}}{\footnotesize}{\small}
 \bibliographystyle{bmc_article}  
  \bibliography{bmc_article} }     

\ifthenelse{\boolean{publ}}{\end{multicols}}{}

\newpage
\section{Appendix}
\subsection{The proof of theorem 1}
  \medskip
  \setcounter{theorem}{0}
  \begin{theorem}
Let $\Sigma$ be an alphabet (say the alphabet of all the amino acids).
For each $l\in\{1,..,L\}$, let $K_l:\Sigma^l\times\Sigma^l\longrightarrow \mathbb{R}$ be a symmetric positive semi-definite kernel. Let $A:\mathbb{R}\longrightarrow \mathbb{R}$ be any function that is the convolution of another function $B:\mathbb{R}\longrightarrow \mathbb{R}$, i.e., for all $z,z'\in\mathbb{R}$, we have
\begin{eqnarray*}
A(z-z')&=&\int_{-\infty}^{+\infty}\, B(z-t) B'(z'-t) \, \mathrm{d}t\,.
\end{eqnarray*}

\bigskip
\noindent
Then, the kernel $K$ defined, for any two strings on the alphabet $\Sigma$, as
\begin{eqnarray*}
K(\xb,\xb')&\eqdef& \sum_{l=1}^L\,\sum_{i=0}^{|\xb|-l}\,\sum_{j=0}^{|\xb'|-l} A(i-j) \ \, K_l\Big((x_{i+1},..,x_{i+l})\,,\,(x'_{j+1},..,x'_{j+l})\Big)
\end{eqnarray*}
is also symmetric positive semi-definite.
\end{theorem}

\medskip
The proof is based on this well known theorem.
\setcounter{theorem}{1}
\begin{theorem}\label{th:Mercer}\cite{cs-00}\\
Let $\Sigma$ be a finite alphabet, 
a string kernel $K:\Sigma^\star\times\Sigma^\star\longrightarrow \mathbb{R}$ is positive semi-definite if and only if it satisfies the Mercer's condition, that is
\begin{eqnarray*}
\sum_{\xb\in\Sigma^\star}\sum_{\xb'\in\Sigma^\star}f(\xb)f(\xb')K(\xb,\xb')&\geq&0
\end{eqnarray*}
for any function $f:\Sigma^\star\longrightarrow \mathbb{R}$ such that $\sum_{\xb\in\Sigma^\star} f^2(\xb)<+\infty$\,.
\end{theorem}

\bigskip
\textbf{Proof of Theorem 1:} \ Clearly, the kernel $K$ is symmetric. So, we only have to show that  it satisfies the Mercer's Condition of Theorem~\ref{th:Mercer}.

For compactness of the notation, the sub-string $(x_{i+1}, x_{i+2}, \ldots, x_{i+l})$ of a string $\xb$ will be denoted $\xb_{]i:i+l]}$.

\bigskip
\bigskip
First note that
\begin{align}
\nonumber
\sum_{\xb\in\Sigma^\star}\sum_{\xb'\in\Sigma^\star}f(\xb)f(\xb')K(\xb,\xb')\ 
 & = \ \ 
\sum_{s=0}^{+\infty}\sum_{s'=0}^{+\infty}\sum_{\xb\in\Sigma^s}\sum_{\xb'\in\Sigma^{s'}} f(\xb)f(\xb')K(\xb,\xb')\\
\nonumber
&=\ \ 
\sum_{s=0}^{+\infty}\sum_{s'=0}^{+\infty}\sum_{\xb\in\Sigma^s}\sum_{\xb'\in\Sigma^{s'}}
   \sum_{l=1}^{L}\sum_{i=0}^{|\xb|-l}\sum_{j=0}^{|\xb'|-l} A(i-j)\  f(\xb)f(\xb')K_l(\xb_{]i,i+l]}, \xb'_{]j,j+l]})\\
   \nonumber
&=\ \ 
\sum_{s=0}^{+\infty}\sum_{s'=0}^{+\infty}\sum_{\xb\in\Sigma^s}\sum_{\xb'\in\Sigma^{s'}}
   \sum_{l=1}^{L}\sum_{i=0}^{s-l}\sum_{j=0}^{s'-l} A(i-j)\  f(\xb)f(\xb')K_l(\xb_{]i,i+l]}, \xb'_{]j,j+l]})\\
   \tag{$\star$}
   & = \ \
   \sum_{l=1}^{L}\sum_{s=0}^{+\infty}\sum_{i=0}^{s-l}\sum_{s'=0}^{+\infty}\sum_{j=0}^{s'-l}A(i-j)\ \sum_{\xb\in\Sigma^s}\sum_{\xb'\in\Sigma^{s'}}
     f(\xb)f(\xb') K_l(\xb_{]i,i+l]}, \xb'_{]j,j+l]})\\
     \nonumber
\end{align}

Note that the last part of the last line of Equation~($\star$) can be rewritten as follows:
\begin{align}
\nonumber
A(i-j) \sum_{\xb\in\Sigma^s}\sum_{\xb'\in\Sigma^{s'}}
     f(\xb)f(\xb') K_l(\xb_{]i,i+l]}, \xb'_{]j,j+l]})\hspace{-70mm}\\
    \nonumber
     & \ \ = \ \ A(i-j)\sum_{\yb\in\Sigma^l}\sum_{\yb'\in\Sigma^{l'}}\ \sum_{\xb\in\Sigma^d\mid\xb_{]i:i+l]}=\yb}\ \sum_{\xb'\in\Sigma^d\mid\xb'_{]i:i+l]}=\yb'}
        f(\xb)f(\xb') K_l(\yb,\yb,)\\
     \tag{$\star\star$}
     &\ 
     \ =\ \ \int_{-\infty}^{+\infty}B(i-t)B(j-t)\, \mathrm{d}t\, \ \ \sum_{\yb\in\Sigma^l}\sum_{\yb'\in\Sigma^{l}}\ \sum_{\xb\in\Sigma^d\mid\xb_{]i:i+l]}=\yb}\ \sum_{\xb'\in\Sigma^d\mid\xb'_{]i:i+l]}=\yb'}
        f(\xb)f(\xb') K_l(\yb,\yb,)\,,
\end{align}
where $\xb\in\Sigma^d\mid\xb_{]i:i+l]}=\yb$ means that  the strings $\xb$ is any string of length $d$ that contains the substring $\yb$ starting at its position $i+1$.
Now, combining $(\star)$ and $(\star\star)$, and putting  
\begin{eqnarray}
g_{t,l}(\yb)&\eqdef &\sum_{s=0}^{+\infty}\sum_{i=0}^{s-l} B(i-t) \sum_{\xb\in\Sigma^d\mid\xb_{]i:i+l]}=\yb} f(\yb)\,,
\end{eqnarray}
for any string $\yb$, any real number $t$ and any $l \in\{1,2,\ldots,L\}$, we  obtain
\begin{align*}
\nonumber
\sum_{\xb\in\Sigma^\star}\sum_{\xb'\in\Sigma^\star}f(\xb)f(\xb')K(\xb,\xb')\ \hspace{-35mm}\\[2mm]
 & = \ \ 
     \sum_{l=1}^L\ \ \sum_{(\yb,\yb')\in \Sigma^l\times\Sigma^l } K_l(\yb,\yb') \sum_{s=0}^{+\infty}\sum_{i=0}^{s-l}\sum_{s'=0}^{+\infty}\sum_{j=0}^{s'-l} \int_{-\infty}^{+\infty}B(i-t)B(j-t)\, \mathrm{d}t\, \hspace{-55mm}\\
     &&\times \left[ \sum_{\xb\in\Sigma^d\mid\xb_{]i:i+l]}=\yb} f(\xb) \right]\ \left[ \sum_{\xb'\in\Sigma^d\mid\xb'_{]j:j+l]}=\yb'} f(\xb') \right]\\[2mm]
 & = \ \ 
    \sum_{l=1}^L\   \int_{-\infty}^{+\infty}\left( \sum_{(\yb,\yb')\in \Sigma^l\times\Sigma^l }K_l(\yb,\yb') \left[  \sum_{s=0}^{+\infty}\sum_{i=0}^{s-l} B(i-t) \sum_{\xb\in\Sigma^d\mid\xb_{]i:i+l]}=\yb} f(\xb)\right]\right.\hspace{-55mm}\\
    &&\times \left.
    \left[  \sum_{s'=0}^{+\infty}\sum_{j=0}^{s'-l} B(j-t) \sum_{\xb'\in\Sigma^d\mid\xb'_{]j:j+l]}=\yb'} f(\xb')\right]
     \right) \, \mathrm{d}t\,\\[2mm]
     & = \ \ 
    \sum_{l=1}^L\   \int_{-\infty}^{+\infty}\left( \sum_{(\yb,\yb')\in \Sigma^l\times\Sigma^l }K_l(\yb,\yb') 
    g_{t,l}(\yb)g_{t,l}(\yb)
     \right) \, \mathrm{d}t\,\\[2mm]
     & \geq 0\,.
\end{align*}
The last line is a consequence of the fact that for every $l$,
$\sum_{(\yb,\yb')\in \Sigma^l\times\Sigma^l }K_l(\yb,\yb') 
    g_{t,l}(\yb)g_{t,l}(\yb)\geq 0$. Indeed, this immediately follows from  Theorem~2, because all the $K_l$'s have been  supposed positive semi-definite. Hence, again by Theorem~2, $K$ itself is also positive semi-definite.
\ebox

\newpage
\subsection{AUC results for experiments on MHC-II}

Calculating the area under the ROC curve (AUC)\cite{Swets1988} for regression problems requires transforming the initial problem into a classification problem \cite{Nielsen2008}. In the case of MHC-peptide binding affinities, the classification problem aims to distinguish between binders and non-binders. To achieve this, a threshold value allowing to distinguish binders from non-binders is required.

\subsubsection{Single-target experiment}
For this experiment, the predicted values were binding energies in kcal/mol. As proposed in \cite{Nielsen2008}, we set a binding affinity threshold of 500nM. Therefore, the threshold value ($t$) in nanomolar was converted to kcal/mol using the technique proposed in \cite{Bordner2010RTA}: \\

$t=-0.586 \times log(500 \times 10^{-9})=8.50207 kcal/mol$.\\

\noindent To calculate the AUC, we converted all the binding energies in the dataset to binary classes based on this binding threshold. Then, for all examples, we predicted the binding energy value ($e$) and generated a confidence value that the given example was a binder. This confidence value is given by: $c = e-t$ and then normalized using all other confidence values to be in the range $[0, 1]$. The latter were used to calculate the AUC for the experiment.

\begin{table}[H]
	\center
	
	\begin{tabular}{lccc}
	\multicolumn{4}{c}{\textbf{AUC}}\\
	\textbf{\textbf{MHC $\boldsymbol{\beta}$ chain}}  & \textbf{KRR+GS} & \textbf{RTA} & \textbf{\# of examples}\\ 
	\hline
	DRB1*0101  & \textbf{0.838} & 0.749 & 5648\\
	DRB1*0301  & \textbf{0.781} & 0.762 & 837\\
	DRB1*0401  & \textbf{0.753} & 0.715 & 1014\\
	DRB1*0404  & 0.786 & \textbf{0.792} & 617\\
	DRB1*0405  & \textbf{0.782} & 0.757 & 642\\
	DRB1*0701  & \textbf{0.845} & 0.790 & 833\\
	DRB1*0802  & 0.724 & \textbf{0.747} & 557\\
	DRB1*0901  & 0.665 & \textbf{0.711} & 551\\
	DRB1*1101  & \textbf{0.830} & 0.753 & 812\\
	DRB1*1302  & 0.708 & \textbf{0.765} & 636\\
	DRB1*1501  & \textbf{0.740} & 0.736 & 879\\
	DRB3*0101  & 0.716 & \textbf{0.825} & 483\\
	DRB4*0101  & \textbf{0.831} & 0.799 & 664\\
	DRB5*0101  & \textbf{0.826} & 0.732 & 835\\
	H2*IA$_b$  & \textbf{0.860} & 0.828 & 526\\
	H2*IA$_d$  & 0.772 & \textbf{0.814} & 306\\
	\hline
	\multicolumn{1}{r}{Average:} & \textbf{0.779} & 0.767 & \\
	\hline
	\end{tabular}
	\caption{Results of the comparison between AUC values for binding energy predictions obtained from Kernel Ridge Regression and the GS Kernel versus the RTA\cite{Bordner2010RTA} method on the dataset proposed by the authors of  this method.}
	\label{resultsHLASingleTargetAUC}
\end{table}

\newpage

\subsubsection{Pan-specific experiment}
The AUC for this experiment was calculated using confidence values as explained above. A threshold of 500 nM was used to discriminate binders from non-binders \cite{Nielsen2008}.

\begin{table}[H]
	\center
	\begin{tabular}{lcccc}
	\multicolumn{5}{c}{\textbf{AUC}}\\
	\textbf{\textbf{MHC $\boldsymbol{\beta}$ chain}}  & \textbf{KRR+GS} & \textbf{MultiRTA} & \textbf{NetMHCIIpan-2.0} & \textbf{\# of examples}\\ 
	\hline
	DRB1*0101   & \textbf{0.807} & 0.801 & 0.794 & 5166\\
	DRB1*0301   & 0.775 & 0.751 & \textbf{0.792} & 1020\\
	DRB1*0401   & \textbf{0.802} & 0.763 & \textbf{0.802} & 1024\\ 
	DRB1*0404   & 0.862 & 0.835 & \textbf{0.869} & 663\\
	DRB1*0405   & \textbf{0.827} & 0.808 & 0.823 & 630\\
	DRB1*0701   & \textbf{0.891} & 0.817 & 0.886 & 853\\
	DRB1*0802   & 0.840 & 0.786 & \textbf{0.869} & 420\\
	DRB1*0901   & \textbf{0.685} & 0.674 & 0.684 & 530\\
	DRB1*1101   & \textbf{0.900} & 0.819 & 0.875 & 950\\
	DRB1*1302   & 0.648 & \textbf{0.698} & 0.648 & 498\\
	DRB1*1501   & \textbf{0.773} & 0.729 & 0.769 & 934\\
	DRB3*0101   & 0.690 & \textbf{0.813} & 0.733 & 549\\
	DRB4*0101   & 0.759 & 0.746 & \textbf{0.762} & 446\\
	DRB5*0101   & \textbf{0.888} & 0.788 & 0.879 & 924\\
	\hline
	\multicolumn{1}{r}{Average:} & 0.796 & 0.773 & \textbf{0.800} & \\
	\hline
	\end{tabular}
	\caption{Results of the comparison between AUC values for binding affinity predictions obtained from Kernel Ridge Regression and the GS Kernel versus the MultiRTA\cite{Bordner2010MultiRTA} and the NetMHCIIpan-2.0\cite{Nielsen2010} methods on the dataset proposed by the authors of NetMHCIIpan\cite{Nielsen2008}.}
	\label{resultsHLAMultiTargetAUC}
\end{table}

\end{bmcformat}
\end{document}